\documentclass[12pt]{iopart}

\usepackage[numbers,sort&compress]{natbib}
\usepackage{graphicx}
\usepackage{epstopdf}
\begin{document}

\title[Graphene antidot lattices: Comparison of Dirac and tight-binding models]{Electronic and optical properties of graphene antidot lattices: Comparison of Dirac and tight-binding models}

\author{S J Brun, M R Thomsen and T G Pedersen}

\address{Department of Physics and Nanotechnology, Aalborg University, Skjernvej 4A, DK-9220 Aalborg \O st, Denmark and\\
Center for Nanostructured Graphene (CNG), DK-9220 Aalborg \O st, Denmark}
\ead{sjb@nano.aau.dk}

\begin{abstract}
The electronic properties of graphene may be changed from semimetallic to semiconducting by introducing perforations (antidots) in a periodic pattern. The properties of such graphene antidot lattices (GALs) have previously been studied using atomistic models, which are very time consuming for large structures. We present a continuum model that uses the Dirac equation (DE) to describe the electronic and optical properties of GALs. The advantages of the Dirac model are that the calculation time does not depend on the size of the structures and that the results are scalable. In addition, an approximation of the band gap using the DE is presented. The Dirac model is compared with nearest-neighbour tight-binding (TB) in order to assess its accuracy. Extended zigzag regions give rise to localized edge states, whereas armchair edges do not. We find that the Dirac model is in quantitative agreement with TB for GALs without edge states, but deviates for antidots with large zigzag regions.
\end{abstract}


\section{Introduction}

Graphene has been the subject of intense research since it was discovered a decade ago \cite{geim04}. This novel two-dimensional material has remarkable electronic \cite{novoselov05,bolotin08}, optical \cite{nair08} and mechanical \cite{lee08} properties. Consequently, it finds potential applications within e.g.\ electronics and optoelectronics \cite{geim09}. The excellent electronic properties of graphene, especially the very high mobility, makes it ideally suited for new smaller and faster nanoelectronic devices \cite{geim04,novoselov05,bolotin08}. Due to its semi-metallic nature, pristine graphene is not well-suited for semiconductor applications. Several strategies for introducing a band gap have been proposed, including graphene nanoribbons (GNRs) \cite{obradovic06,son06,han07}, gated bilayer graphene \cite{zhang09} and periodic gating \cite{pedersen12band}. Another method is to introduce perforations in a periodic pattern, called a graphene antidot lattice (GAL) \cite{pedersen08a,pedersen08}. This provides a controllable band gap that depends on the geometry of the antidot lattice \cite{pedersen08a}. Previously, tight-binding (TB) calculations have been made for relatively small unit cells \cite{pedersen08a,pedersen08,furst2009}. Trolle \textit{et al.}\ \cite{trolle2013large} have used density functional theory (DFT) and Hubbard TB to show that localized edge states emerge in GALs containing hexagonal antidots with zigzag edges. However, realistic structures are typically much larger than the ones studied theoretically, and the calculation time scales badly with the size of the structures. F\"urst \textit{et al.}\ \cite{furst2009} have previously presented an analysis based on the Dirac equation (DE), in which they used finite-element analysis to calculate the electronic properties of GALs with circular antidots. The computational time of their method depends only on the ratio between the radius of the antidot and the size of the unit cell, but their method only qualitatively predicts the band structure.

Recently, GALs with circular antidots have been fabricated by several groups \cite{eroms09,giesbers12,kim10,kim12}. Such structures are fabricated either by e-beam lithography \cite{eroms09,giesbers12} or using diblock copolymer templates \cite{kim10,kim12}. Moreover, Oberhuber \textit{et al.}\ \cite{oberhuber2013weak} have fabricated GALs with hexagonal antidots. They used an etching technique that selectively etches armchair edges, which produces hexagonal antidots with zigzag edges. Xu \textit{et al.}\ \cite{xu2013controllable} have demonstrated that it is possible to create antidots with diameters down to 2 nm using a scanning transmission electron microscope. When subsequently heating the sample, the curved edges of the antidots were observed to reconstruct into armchair edges. It has also been shown that Joule heating reconstructs graphene edges into zigzag or armchair configurations \cite{jia2009controlled}. Theoretical studies based on DFT show that the preferred edge chirality of GNRs is armchair in an oxygen-rich atmosphere and zigzag for water-saturated GNRs \cite{seitsonen2010structure}. Although there may still remain some edge roughness, these findings show that the chirality of the edges of GNRs and GALs is controllable. 

In this paper, we present a continuum model of GALs based on the DE. In this method, the antidot lattice is modelled by a spatially varying mass term that is only nonzero inside the antidots. This makes the antidot regions increasingly unfavourable for electrons as the mass term increases. The major advantage of the Dirac model is that the calculation time does not depend on the size of the structure that is being studied. In fact, for energies much smaller than the mass term, the results are scalable. This means that, e.g., a given band structure can be used to describe a geometry where all lengths are scaled by some factor if the energies are divided by the same factor. The Dirac model is compared with nearest-neighbour TB in order to assess its accuracy. The two models will mainly be compared for GALs containing hexagonal antidots with zigzag or armchair edges. Furthermore, the DE is used to derive an approximation of the band gap of GALs, which is compared with TB for a wide range of structures. We demonstrate that the Dirac model is in quantitative agreement with TB for GALs containing antidots with armchair edges. However, for other antidot geometries, the models only agree for small antidots.

\section{Theory and methods}

\begin{figure}[tb]
\centering
\includegraphics[width=8.5cm]{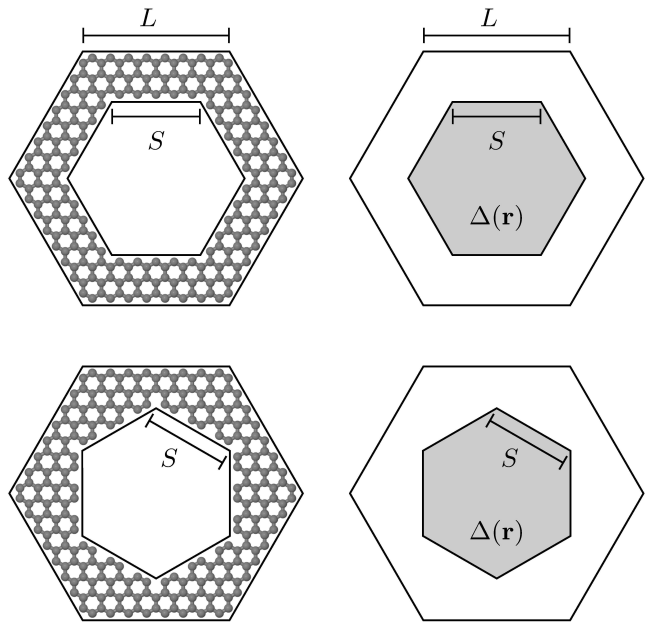}
\caption{Unit cells used in TB (left) and DE (right) for hexagonal antidots with zigzag (top) and armchair (bottom) edges in triangular antidot lattices. The atomic structures shown are Z$\{8,5\}$GAL and A$\{8,5\}$GAL.}
\label{fig:geometry}
\end{figure}

In the present work, we will model GALs using the DE and compare the results with nearest-neighbour TB. We use the notation GAL to describe structures where the antidot lattice vectors are parallel to the carbon-carbon bonds. By rotating the lattice $\pi/6$, the antidot lattice vectors are perpendicular to the carbon-carbon bonds. These structures will be denoted rotated GALs (RGALs) as in \cite{petersen2010clar}. We will focus on GALs containing hexagonal antidots with zigzag and armchair edges, which we will refer to as zigzag and armchair antidots throughout the paper. Figure~\ref{fig:geometry} shows examples of GALs with zigzag and armchair antidots used in TB and the Dirac model. GALs with circular antidots and RGALs with armchair antidots will also be considered. The structures are described by the side length $L$ of the unit cell and the side length $S$ of the antidot, where all distances are in units of the graphene lattice constant~$a$. Circular antidots are correspondingly characterized by the radius $R$. The unit cells for TB are generated by removing all atoms within the antidot region and subsequently removing dangling bonds. The notations Z$\{L,S\}$GAL and A$\{L,S\}$GAL will be used to describe the geometry of GALs with zigzag and armchair antidots, respectively. Furthermore, the notations C$\{L,R\}$GAL and A$\{L,S\}$RGAL will describe GALs with circular antidots and RGALs with armchair antidots, respectively.

The Dirac Hamiltonian for a graphene lattice with a spatially varying mass term $\Delta(\mathbf{r})$ has the form \cite{furst2009}

\begin{equation}
\mathbf{H} = \left(\begin{array}{cc}
	\Delta(\mathbf{r}) & -\hbar v_F\left( i\partial_x-\partial_y \right) \\
	-\hbar v_F\left( i\partial_x+\partial_y \right) & -\Delta(\mathbf{r})
\end{array}\right),
\end{equation}

\noindent where the mass term has a constant value of $\Delta_0$ inside the antidot and is vanishing elsewhere. The wave function $\Psi$ will satisfy the Bloch condition if $\Psi(\mathbf{r}) = e^{i\mathbf{k}\cdot\mathbf{r}}u(\mathbf{r})$, where the function $u(\mathbf{r})$ is a lattice-periodic spinor containing the components $u^A(\mathbf{r})$ and $u^B(\mathbf{r})$. We then express $\Delta(\mathbf{r})$ and $u(\mathbf{r})$ as Fourier series, as they are both periodic with the antidot lattice

\begin{equation}
\Delta(\mathbf{r}) = \sum_{\mathbf{G}}{\Delta_\mathbf{G}e^{i\mathbf{G}\cdot\mathbf{r}}}, \quad
u(\mathbf{r}) = \sum_{\mathbf{G}}{u_\mathbf{G}e^{i\mathbf{G}\cdot\mathbf{r}}},
\end{equation}

\noindent where $u_\mathbf{G}$ is a spinor containing the Fourier coefficients $u^A_\mathbf{G}$ and $u^B_\mathbf{G}$, $\mathbf{G} = p\mathbf{g}_1+q\mathbf{g}_2$ is the reciprocal lattice vector, $p$ and $q$ are integers, and $\mathbf{g}_1$ and $\mathbf{g}_2$ are the primitive reciprocal lattice vectors of the antidot lattice. The geometry of the antidot is then solely described by the Fourier coefficients $\Delta_\mathbf{G}$ and the geometry of the unit cell is solely described by $\mathbf{g}_1$ and $\mathbf{g}_2$. The expression for $\Delta_\mathbf{G}$ for an arbitrary $N$-sided polygon was derived in \cite{lee83fourier}. Inserting the expressions for $\Delta(\mathbf{r})$ and $\Psi(\mathbf{r})$ in the Dirac equation $\mathbf{H}\Psi=E\Psi$ leads to the expression

\begin{equation}
\sum_\mathbf{G'}{\mathbf{H}_{\mathbf{G},\mathbf{G'}} u_\mathbf{G'}} = Eu_\mathbf{G},
\end{equation}
\begin{equation}
\mathbf{H}_{\mathbf{G},\mathbf{G'}} = 
\left(\begin{array}{cc}
	\Delta_{\mathbf{G}-\mathbf{G'}} & T_\mathbf{G}\delta_{\mathbf{G},\mathbf{G'}} \\ 
	T^*_\mathbf{G}\delta_{\mathbf{G},\mathbf{G'}} & -\Delta_{\mathbf{G}-\mathbf{G'}}
\end{array}\right),
\end{equation}

\noindent where $T_{\mathbf{G}} = \hbar v_F [k_x+G_x-i(k_y+G_y)]$. This may be set up as a matrix equation and solved as an eigenvalue problem through numerical diagonalization. 
Electrons are excluded more and more from the antidot region as the mass term increases, and in the limit of an infinite mass term, the electrons are completely excluded. Therefore, convergence is obtained by using a sufficiently large mass term. 
However, convergence must also be ensured by choosing a basis that is large enough. Throughout the paper we use a mass term given by $\Delta_0=170$~eV$/L$. The reciprocal lattice vectors used for the basis are created by letting $p,q\in[-N,N]$, where we use $N=20$ and $N=16$ for hexagonal and circular antidots, respectively. These parameters were found to provide adequately converged results.

Our method is different from the one used by F\"urst \textit{et al.}\ \cite{furst2009}, who studied GALs with circular antidots using the DE. They used the commercially available finite-element solver COMSOL Multiphysics for their calculations. They studied the case of an infinite mass term by imposing the boundary condition that the current normal to the edge of the antidot is vanishing. This method was shown to provide results that agree qualitatively, but not quantitatively, with TB.
Their boundary condition states that $\Psi_A(\mathbf{r})=ie^{-i\phi}\Psi_B(\mathbf{r})$, where $\Psi_{A/B}(\mathbf{r})$ are the two spinor components of the wave function and $\phi$ is the polar angle of the normal vector at a given point on the edge of the antidot. This was shown to be problematic in the limit of vanishing antidots where the angle $\phi$ becomes completely undetermined. In this case, the band gap was non-vanishing and approached a value of approximately $1.02\gamma/L$, where $\gamma$ is the transfer integral of nearest-neighbour TB. 
Our method uses a finite mass term. However, in the limit of an infinite mass term, the two approaches should be equivalent, and in this case our method should also show a finite band gap in the limit of vanishing antidots. In practice, we cannot use an infinite mass term, as this would require an infinite basis. Because our model uses a finite mass term, we do not encounter the same problem in the limit of vanishing antidots. 

We have focused our attention on hexagonal antidots, although other geometries may easily be considered by adjusting the Fourier coefficients of the mass term accordingly. An approximation of the band gap of a GAL is derived in \ref{app:appendix} from the DE by assuming cylindrical symmetry in the unit cell. 

The atomistic model used for comparison is nearest-neighbour TB in the orthogonal approximation (assuming no overlap between atomic wave functions) with a transfer integral of $\gamma=3.033$~eV.

\section{Results}
In this section, we present the results of our Dirac model and compare them with TB. Only positive energies of band structures will be shown, as the valence bands follow from exact electron-hole symmetry. We will present results for GALs with zigzag and armchair antidots as well as GALs with circular antidots and RGALs with armchair antidots.

\begin{figure}[tb]
\centering
\includegraphics[width=8.5cm]{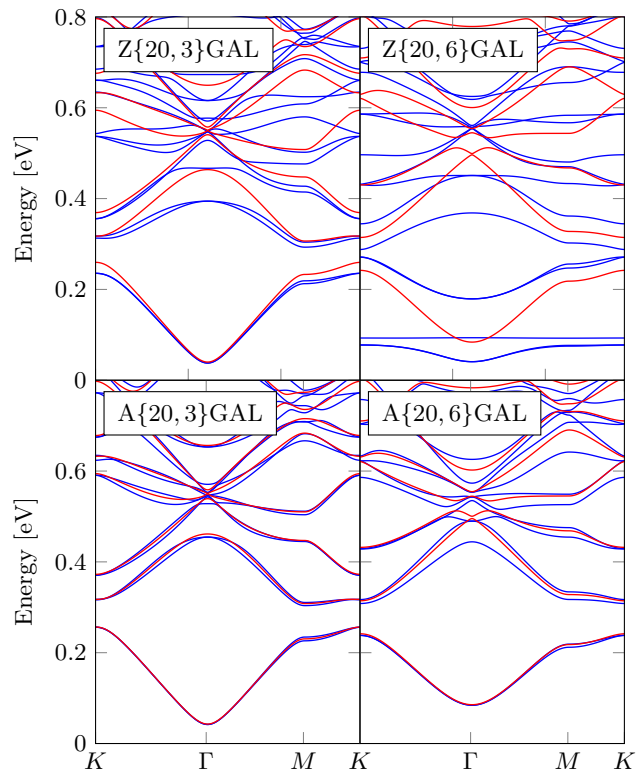}
\caption{Comparison of DE (red) and TB (blue) band structures for GALs with zigzag and armchair antidots.}
\label{fig:band_structures}
\end{figure}

Band structures calculated using the DE and TB are compared in figure~\ref{fig:band_structures} for four different geometries. The geometries used for the Dirac model are created such that the area of the antidot equals the total area of the removed atoms. For all four geometries shown, a band gap opens up at the $\Gamma$-point both for the DE and TB calculations. In the case of zigzag antidots, the Dirac model agrees well with the band structure from TB when the antidot is very small, e.g.\ for the Z$\{20,3\}$GAL geometry. However, large discrepancies are observed for the Z$\{20,6\}$GAL geometry. The band structures agree much better for armchair antidots. For the A$\{20,3\}$GAL geometry, the DE band structure almost coincides with the TB band structure, and the two models are in excellent agreement in this case. Even the band structures for the A$\{20,6\}$GAL with a larger antidot agree very well. This tendency continues for larger antidots, where the band structures from the two models remain very similar. 

\begin{figure}[tb]
\centering
\includegraphics[width=7.1cm]{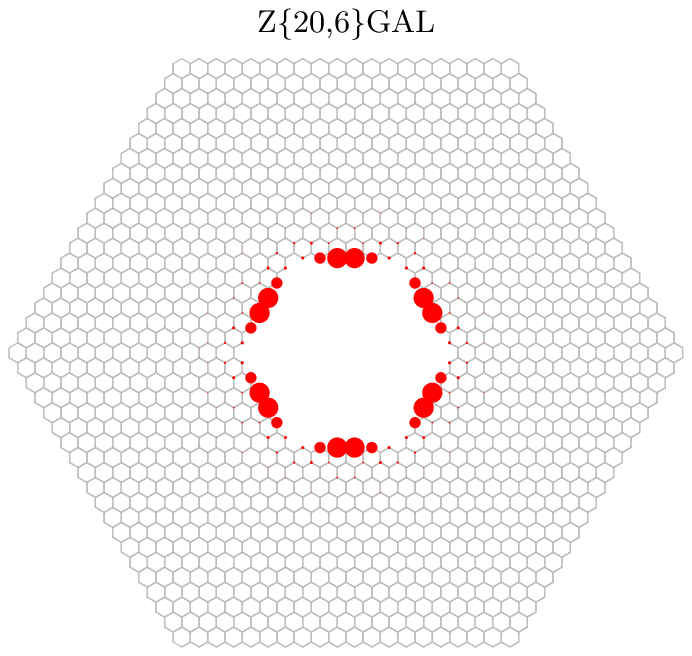}\\
\includegraphics[width=7.1cm]{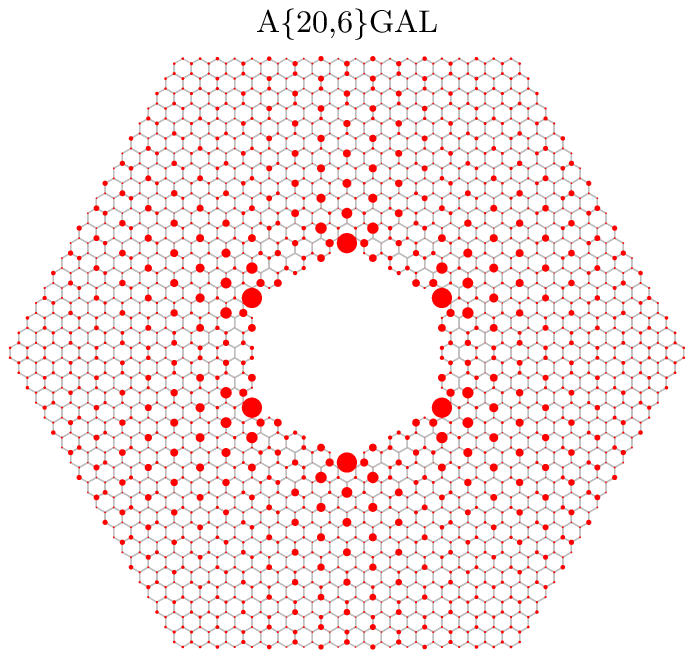}
\caption{Electron probability density of the third conduction band in one unit cell. The radius of each circle is proportional to the absolute square of the eigenvector element for that atom and chosen such that the radius of the largest circle is the same in both plots.}
\label{fig:probdens}
\end{figure}

The lowest bands of the Z$\{20,6\}$GAL geometry are very flat, especially the third conduction band near 0.09~eV, which is almost completely dispersionless. For larger zigzag antidots, even more bands become dispersionless, and the band structures agree even worse. Dispersionless bands are associated with localized states. The localization of the electrons may be visualized by plotting the electron probability density on each atom in the unit cell. Figure~\ref{fig:probdens} shows the probability density of the third conduction band for the Z$\{20,6\}$GAL and A$\{20,6\}$GAL geometries within one unit cell. The plots are generated by averaging over the Brillouin zone. 
It is clear that the electrons of the Z$\{20,6\}$GAL are confined to the edge of the antidot, whereas the electrons in the A$\{20,6\}$GAL are generally spread out over the entire unit cell and only slightly localized in the corners of the antidot. Such localized edge states are generally observed when the antidot contains long zigzag regions. The existence of localized edge states was studied by Fujita \textit{et al.}\ \cite{fujita1996peculiar}, who showed that edge states appear for semi-infinite graphene with zigzag termination, whereas armchair termination does not lead to edge states. 
Brey and Fertig \cite{brey2006electronic} have used the DE to study the electronic states of GNRs, and by using appropriate boundary conditions, they arrived at the same conclusion. 
Localized edge states in GALs have previously been studied by Vanevi\'c \textit{et al.}\ \cite{vanevic2009character}. They showed that triangular antidots with zigzag edges lead to dispersionless bands where the electrons are localized at the edge of the antidot, which is in good agreement with our results. 
Recently, Trolle \textit{et al.}\ \cite{trolle2013large} used DFT and Hubbard TB to investigate localized edge states in GALs with zigzag antidots. Furthermore, they showed that the edge states become spin polarized when $S\geq6$. 
Edge states have also been observed experimentally using scanning tunnelling spectroscopy on GNRs fabricated by "unzipping" carbon nanotubes \cite{tao2011spatially}. Edge states modify the electronic properties of GNRs and figure~\ref{fig:band_structures} shows that they also modify the electronic properties of GALs. As the size of the antidot increases, edge states appear for zigzag antidots and the electrons become more and more confined to the edges of the antidot. 
The Dirac model is a continuum model, and consequently all atomistic features are missing. With no boundary conditions, the Dirac model is unable to predict the localized edge states appearing for zigzag edges.

\begin{figure}[tb]
\centering
\includegraphics[width=8.5cm]{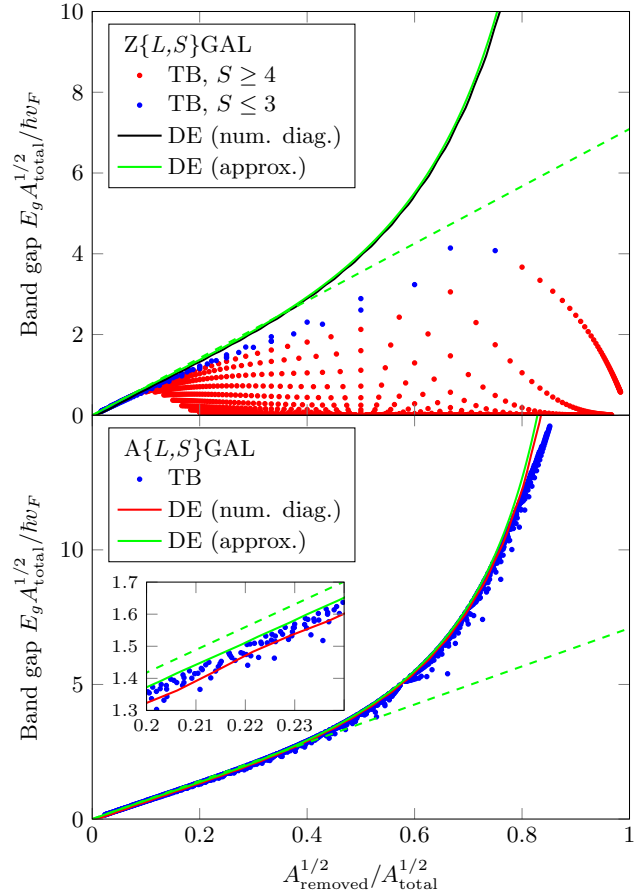}
\caption{Band gap of GALs with zigzag and armchair antidots calculated using TB and the DE. The green dashed line is a linearisation of the approximated Dirac curve. The inset shows a zoom of the linear region.}
\label{fig:bandgap_GAL}
\end{figure}

The size of the band gap is highly dependent on the lattice geometry. Generally, the band gap increases as the ratio of antidot to unit-cell area (fill factor) increases. 
A linear scaling law for GALs with circular antidots was proposed by Pedersen \textit{et al.}\ \cite{pedersen08a} suggesting that the band gap scales as $E_g \approx K\cdot N_{\mbox{\scriptsize{removed}}}^{1/2}/N_{\mbox{\scriptsize{total}}}$ for small values of $N_{\mbox{\scriptsize{removed}}}^{1/2}/N_{\mbox{\scriptsize{total}}}$, where $N_{\mbox{\scriptsize{removed}}}$ is the number of removed atoms and $N_{\mbox{\scriptsize{total}}}$ is the total number of atoms in the unit cell before the antidot was created. They determined the scaling constant as $K\simeq25$~eV, whereas a more exact quasiparticle TB model has revealed a slightly larger constant of $K\simeq29$~eV \cite{petersen2009quasiparticle}. The DE band structures in figure~\ref{fig:band_structures} show that the band gap increases as the size of the antidot increases, which is expected from the scaling law. 
The size of the band gap may be estimated by replacing the hexagonal unit cell with an approximated unit cell with full cylindrical symmetry and by assuming an infinite mass term. This means that both the unit cell and the antidot are replaced by circles of equivalent areas, see \ref{app:appendix} for a derivation. The band gap then only depends on the total area of the unit cell $A_{\mbox{\scriptsize{total}}}$ and the area of the antidot $A_{\mbox{\scriptsize{removed}}}$. 
The approximation of the band gap (given by \ref{eq:analytic1}) may be used to calculate the band gap scaled by $A_{\mbox{\scriptsize{total}}}^{1/2}$ as a function of $A_{\mbox{\scriptsize{removed}}}^{1/2}/A_{\mbox{\scriptsize{total}}}^{1/2}$, which becomes the universal curve shown in figure~\ref{fig:bandgap_GAL}. 
The scaling law predicts a linear correlation on these axes, and a linear approximation of \ref{eq:analytic1} (given by \ref{eq:analytic2}) is also shown in the figure. 
The scaling constant for the DE, obtained from \ref{eq:analytic2}, is $K = 4\cdot3^{1/4}\sqrt{\pi}\gamma \simeq 28.3$~eV, which is very close to the scaling constants determined from atomistic models. 

Band gap energies of a wide range of structures have been calculated using TB and are compared with the results of the Dirac model in figure~\ref{fig:bandgap_GAL}. The approximation of the band gap using the DE is also included in the figure. 
The values of $A_{\mbox{\scriptsize{total}}}$ and $A_{\mbox{\scriptsize{removed}}}$ in TB are calculated directly from $N_{\mbox{\scriptsize{total}}}$ and $N_{\mbox{\scriptsize{removed}}}$, respectively. 
The approximated band gap is seen to be a very good estimate as it is very close to the curve obtained from the numerical diagonalization method. Furthermore, the Dirac model predicts that the band gap increases linearly in the regime $A_{\mbox{\scriptsize{removed}}}^{1/2}/A_{\mbox{\scriptsize{total}}}^{1/2}<0.4$. 
For GALs with zigzag antidots, the TB results are close to the results from the Dirac model when the antidots are fairly small. However, edge states appear for larger antidots, which cause the band gap to shrink. Furthermore, the band gaps from TB for zigzag antidots are always lower than the linear Dirac result. 
For the A$\{L,S\}$GAL structures, the band gaps calculated from TB are all very close to the curves from the DE. Moderate deviations are only observed in the region $A_{\mbox{\scriptsize{removed}}}^{1/2}/A_{\mbox{\scriptsize{total}}}^{1/2}>0.8$. The absence of localized edge states in the case of armchair edges means that the band gap does not vanish for large antidots. The inset in the figure shows a zoom, where it is seen that the approximation of the band gap from the DE serves as an upper limit for the TB band gap calculations. 
The band gap obtained from the numerical diagonalization method shows lower values than the approximated version. This is partly because the numerical diagonalization uses a finite mass term, and partly because the approximated band gap is calculated using an approximate geometry (assumes cylindrical symmetry). 
Clearly, the results of figure~\ref{fig:bandgap_GAL} show that the DE is able to accurately predict the band gap of GALs with armchair antidots.

\begin{figure}[tb]
\centering
\includegraphics[width=8.5cm]{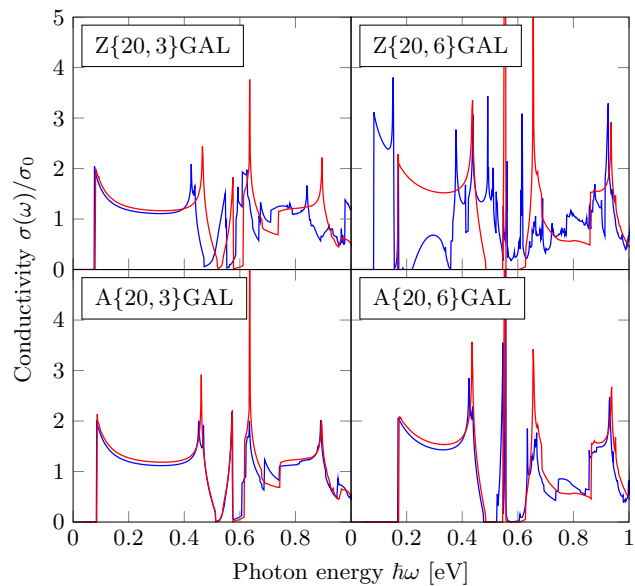}
\caption{Optical conductivity in units of the pristine graphene conductivity $\sigma_0=e^2/4\hbar$ calculated using the DE (red) and TB (blue) for the GALs shown in figure~\ref{fig:band_structures}.}
\label{fig:intercond}
\end{figure}

The approximation of the band gap using the DE seems to be the better choice, as it is computationally much faster than numerical diagonalization. 
However, the numerical diagonalization method is necessary in order to calculate band structures and may also be used to calculate other properties such as the density of states and optical conductivity. A comparison of the optical conductivity calculated using the DE and TB is shown in figure~\ref{fig:intercond} for four GALs. The method for calculating the optical conductivity was adopted from \cite{pedersen08}. We reach the same conclusion as for the band structures in figure~\ref{fig:band_structures}. The optical conductivity from the Dirac model agrees very well with the TB results for armchair antidots. For zigzag antidots, the results agree for low energies when the antidot is small, but the optical spectra are very different for larger antidots, e.g.\ the Z$\{20,6\}$GAL. The optical properties of gapped graphene, i.e.\ using a spatially invariant mass term, have previously been presented in a closed-form expression and compared with TB \cite{pedersen09}. The conductivity spectra $\sigma(\omega)$ were shown to always increase abruptly at the band gap energy to $\sigma(\omega_g)=2\sigma_0$, where $\sigma_0=e^2/4\hbar$ is the conductivity of pristine graphene. Gapped graphene was shown to be a good approximation at energies near the band gap for a GAL with a small circular antidot. The spectra from our Dirac model follow the spectra from TB very well in the case of armchair antidots, and even capture features at energies far from the band gap. 

\begin{figure}[tb]
\centering
\includegraphics[width=8.5cm]{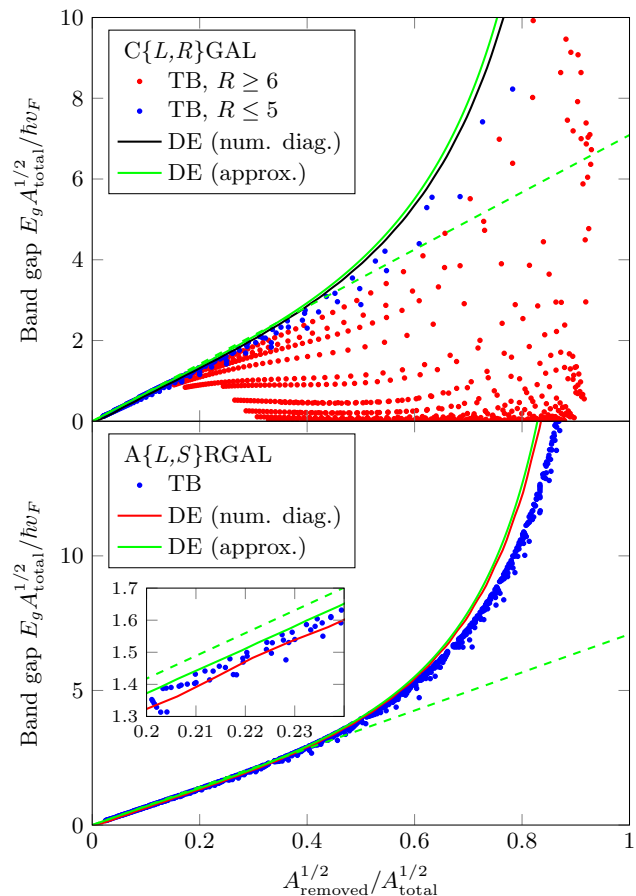}
\caption{Same as figure~\ref{fig:bandgap_GAL}, but for GALs with circular antidots and RGALs with armchair antidots.}
\label{fig:bandgap_RGAL}
\end{figure}

Until now, we have only considered GALs with hexagonal antidots, but other geometries may easily be compared with the Dirac model. 
Figure~\ref{fig:bandgap_RGAL} shows a comparison of the band gap calculated using the DE and TB for GALs with circular antidots and RGALs with armchair antidots. The edge of circular antidots will consist of both zigzag and armchair edges when the antidot is not very small. The zigzag parts of the edge will support localized edge states when the antidot is large, which cause the band gap to shrink as observed in the figure. However, for small antidots ($R\leq 5$), the Dirac model predicts the band gap reasonably well, as the localization is weak. 

The band gap calculations of A$\{L,S\}$RGAL structures show the same tendency as the A$\{L,S\}$GAL structures in figure~\ref{fig:bandgap_GAL}. RGALs were found to provide a band gap only for every third value of $L$. This is consistent with previous findings \cite{petersen2010clar}, and also obeys a universal band gap opening rule by Dvorak \textit{et al.}\ \cite{dvorak2013bandgap}. The Dirac model predicts that the band gap increases dramatically for $A_{\mbox{\scriptsize{removed}}}^{1/2}/A_{\mbox{\scriptsize{total}}}^{1/2}>0.8$. TB shows somewhat lower values of the band gap in this region, but these also increase dramatically as for the Dirac model. 
Again, the inset shows that the approximation of the band gap from the DE seems to be the upper limit of TB.	

It should be noted that while all structures considered in this paper are perfectly ordered, realistic structures from experiments will to some extent contain disorder. Theoretical studies have shown that the band gap of GALs is robust against a considerable amount of disorder \cite{yuan2013electronic}. The band gap was found to initially shrink and eventually vanish as the amount of disorder increased. Other calculations have shown that the properties of graphene waveguide structures based on GALs are also robust against structural disorder \cite{pedersen2012graphene}.

We have shown that the Dirac model is in good agreement with TB in the absence of edge states. However, in case of zigzag or circular antidots, edge states cause the band gap to shrink. 
If the electrons of edge states are completely confined to the edges of the antidot, they will not be able to contribute to the electronic transport of the GAL. 
The lowest conduction bands of GALs with large zigzag antidots are almost completely dispersionless, which suggests that the transport gap in such cases may be larger than the band gap.

\section{Conclusion}

We have presented a continuum model based on the Dirac equation, which describes the electronic and optical properties of graphene antidot lattices. The major advantages of the Dirac model are that the computational time does not depend on the size or geometry of the structures, and that the results are scalable. The Dirac model is compared with tight-binding calculations of the corresponding atomistic structures in order to determine its accuracy. A comparison of band structures shows that the Dirac model is in quantitative agreement with tight-binding for structures with no edge states, e.g.\ antidots with armchair edges. 
The present Dirac model is unable to predict edge states as it does not distinguish between zigzag and armchair edges. 
Comparing band gap calculations and optical spectra also shows quantitative agreement between the models for structures with no edge states.

An approximation of the band gap of a graphene antidot lattice was derived from the Dirac equation. A linearisation revealed a scaling constant in good agreement with previously suggested values obtained from atomistic models. 
The approximation provides a very fast way of estimating the band gap of a graphene antidot lattice with no edge states even if the antidot makes up a large part of the unit cell.

\section*{Acknowledgments}
The authors gratefully acknowledge the financial support from the Center for Nanostructured Graphene (Project No. DNRF58) financed by the Danish National Research Foundation. We thank A.-P. Jauho for useful comments on the manuscript.

\appendix
\section{Estimate of band gap}
\label{app:appendix}

\begin{figure}[htb]
\centering
\includegraphics[width=8cm]{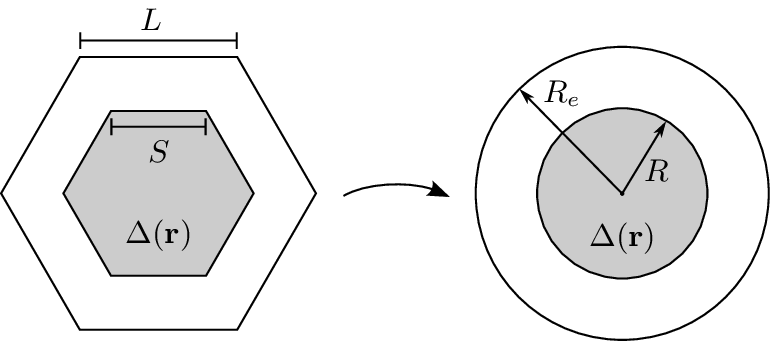}
\caption{The hexagonal unit cell and the antidot are replaced by circles of equivalent areas. This approximated geometry has cylindrical symmetry.}
\label{fig:appendix}
\end{figure}

In this appendix, we present an approximation of the band gap of GALs derived using the DE. The hexagonal unit cell is replaced by one with full cylindrical symmetry, i.e.\ a circle of radius $R_e$, see figure~\ref{fig:appendix}. 
This approach is inspired by \cite{mortensen2006}. The area of the circle is equal to the area of the hexagonal unit cell, such that $A_{\mbox{\scriptsize{total}}}=\pi R_e^2$. If the antidot is not circular, this is also replaced by a circle with radius $R$ of equivalent antidot area, $A_{\mbox{\scriptsize{removed}}}=\pi R^2$. 

The Dirac Hamiltonian in cylindrical coordinates is

\begin{equation}
\mathbf{H} = \hbar v_F\left(\begin{array}{cc}
\widetilde{\Delta}(r) & -i e^{-i\theta}\left(\partial_r-\frac{i}{r}\partial_\theta\right)\vspace{1.5mm}\\
-i e^{i\theta}\left(\partial_r+\frac{i}{r}\partial_\theta\right) & -\widetilde{\Delta}(r)
\end{array}\right),
\end{equation}

\noindent where $\widetilde{\Delta}(r) = \frac{\Delta_0}{\hbar v_F} H(R-r) = \widetilde{\Delta}_0 H(R-r)$ and $H$ is the Heaviside step function. The wave function is of the form

\begin{equation}
\Psi(r,\theta) = \frac{1}{\sqrt{2}} \left( \begin{array}{c}
 i^mf(r)e^{im\theta} \\
i^{m+1}g(r)e^{i(m+1)\theta} \end{array} \right),
\end{equation}

\noindent which is inserted in the DE together with the Hamiltonian. For a piecewise constant mass term, the solutions for $f$ and $g$ are 

\begin{equation}
f(r) = \left\{ \begin{array}{ll}
	J_m(kr)+B_mY_m(kr) & \enskip r>R\\
	C_mI_m(qr) & \enskip r<R
\end{array}\right.,
\end{equation}

\begin{equation}
g(r) = \left\{ \begin{array}{ll}
	J_{m+1}(kr)+B_mY_{m+1}(kr) & \enskip r>R\\
	-C_m\sqrt{\frac{\widetilde{\Delta}_0-k}{\widetilde{\Delta}_0+k}}I_{m+1}(qr) & \enskip r<R
\end{array}\right.,
\end{equation}

\noindent where $J_m$ and $Y_m$ are the $m$'th order Bessel functions of the first and second kind, respectively, $I_m$ is the $m$'th order modified Bessel function of the first kind, $k = E/\hbar v_F$ and $q = (\widetilde{\Delta}_0^2-k^2)^{1/2}$. Both $f$ and $g$ must be continuous at $r=R$, which is used to determine $B_m$ and $C_m$. For the lowest state ($m=0$) and in the limit of large $\widetilde{\Delta}_0$, the coefficients become 

\begin{equation}
B_0 \approx -\frac{J_0(kR)+J_1(kR)}{Y_0(kR)+Y_1(kR)}, \quad C_0\approx 0.
\end{equation}

\noindent This derivation is generally carried out using a finite mass term, and the band gap may also be calculated in this case. The wave functions inside and outside the antidot are matched at the edge of the antidot in the case of a finite mass term, which is used to determine $B_m$ and $C_m$. Subsequently the limit of a large mass term is applied for which the coefficients listed above are valid. This approach does not lead to boundary conditions that cause problems in the limit of small antidots as observed in \cite{furst2009}. 

We restrict our analysis to the $\Gamma$-point of the Brillouin zone, as this is where the band gap opens. 
We still require that the wave function is Bloch-periodic when using the approximated geometry. 
However, at the $\Gamma$-point it is merely periodic. Periodicity implies a vanishing derivative of $f$ at the outer boundary ($r=R_e$), meaning that $J_1(kR_e)+B_0Y_1(kR_e)=0$. This yields the equation 

\begin{equation}
J_1(kR_e)[ Y_0(kR)+Y_1(kR) ] - Y_1(kR_e)[ J_0(kR)+J_1(kR) ] = 0,
\label{eq:analytic1}
\end{equation}

\noindent which may be solved numerically for $k$ to obtain the band gap given by $E_g=2\hbar v_F k$. $f$ is used to solve for the lowest energy of the conduction bands. Equivalently, $g$ may be solved for negative energies using $m=-1$ which leads to the highest energy of the valence bands. The Bessel functions in \ref{eq:analytic1} are approximated by assuming small $k$, such that the equation becomes

\begin{equation}
\frac{4}{kR_e}+kR_e-\frac{2R_e}{R}+\frac{2R}{R_e}+kR_e(2+kR)\ln\left( \frac{R}{R_e} \right)=0.
\end{equation}

\noindent In the limit of small $R$, the solution becomes the simple expression $k\approx 2R/R_e^2$, meaning that

\begin{equation}
E_g = 4\hbar v_F \frac{R}{R_e^2}=4\sqrt{\pi}\hbar v_F \frac{A_{\mbox{\scriptsize{removed}}}^{1/2}}{A_{\mbox{\scriptsize{total}}}}.
\label{eq:analytic2}
\end{equation}

\noindent This shows that at small $k$, the band gap is directly proportional to the square root of the removed area and inversely proportional to the area of the unit cell, which is consistent with previously suggested scaling laws \cite{pedersen08a,petersen2009quasiparticle}.

\bibliography{literature}

\end{document}